\documentclass[12pt]{article}
\usepackage{times}
\usepackage{geometry}
\usepackage[utf8]{inputenc}
\usepackage{enumitem,amssymb}
\usepackage{graphicx}
\usepackage{authblk}
\usepackage{hyperref}
\usepackage{natbib}

\title{Robust Archives Maximize Scientific Accessibility}

\begin{document}
\author[1,2]{J.E.G. Peek}
\author[3]{Vandana Desai}
\author[1]{Richard L. White}
\author[4]{Raffaele D’Abrusco}
\author[3]{Joseph M. Mazzarella}
\author[4]{Carolyn Grant}
\author[1]{Jenny L. Novacescu}
\author[3]{Elena Scire}
\author[4]{Sherry Winkelman}

\affil[1]{Space Telescope Science Institute, 
3700 San Martin Drive, 
Baltimore, MD, 21218}

\affil[2]{Department of Physics \& Astronomy, 
Johns Hopkins University,
3400 N. Charles Street, 
Baltimore, MD 21218}

\affil[3]{Infrared Processing and Analysis Center, 
California Institute of Technology, 
MS 100-22, 
Pasadena, 
CA 91125, 
USA}

\affil[4]{Smithsonian Astrophysical Observatory, 
60 Garden Street, 
02138 Cambridge, MA, USA}

\maketitle

\section{Introduction}

\emph{The High Impact of Astronomical Data Archives} \citep{White2009} demonstrated that astronomical archives are very important for astronomy’s scientific productivity. Specifically, it was shown that for each paper written using data from \emph{Chandra} and \emph{Hubble} by someone who proposed the observation (``guest observer''), there is at least one written by someone who was not on the original proposal. Further, it was shown that these ``archival'' papers are just as common among well cited mission publications as the non-archival papers.

Scientific productivity is a key metric of success for astronomy, but it is not the only one. This contribution updates and expands the results from \citet{White2009}, but primarily focuses on scientific \emph{accessibility}. If we measure the scientific productivity of a mission or telescope by the number of publications per unit time, we can measure the scientific \emph{accessibility} of a mission as the number of different authors, institutions, communities, or countries publishing using those data. The scientific accessibility of astronomical data is critical to maintain a rich, flourishing, and growing discourse in astronomy. If the astronomy conversation is dominated by only a few voices, institutions, or countries, the entire scientific process, where old ideas are constantly challenged and new ideas are constantly proposed, can wither and die. Further, by expanding the community working on these missions and in astronomy we sow the seeds for the future success of the discipline. We note that engagement of the lay community through public outreach and citizen science is also critical to the success of astronomy and is similarly enhanced by access to archival data, but in this work we explicitly address scientific engagement with astronomical data. 

In this work we use a bibliographic analysis to show that the \emph{Chandra}, \emph{Hubble}, and \emph{Spitzer} archives remain very scientifically productive and dramatically enhance scientific accessibility. We then make recommendations based on these findings.

\section{Data \& Methodology}
Our analysis of scientific accessibility (and update on scientific productivity) for \emph{Chandra}, \emph{Hubble}, and \emph{Spitzer} begins with their respective telescope bibliographies. All publications that use data from each of these missions are recorded by bibliographers and assigned the tag “guest observer”, “archival” or “both”. While details differ, \emph{Hubble} and \emph{Spitzer} bibliographies use very similar rules to sort papers into these categories. If a data set was used and no author is a principal or co- investigator on the original proposal under which these data were taken, the proposal is marked “archival”. If at least one author is a principal or co-investigator, it is marked “guest observer”. Finally, if data from more than one program is used, and the status is mixed, it is marked “both”. \emph{Chandra} uses a slightly different definition of archival, in which a publication is “archival” if the PI and prime observer are not authors on the publication using that data. In the vast majority of cases, this bibliographic technique would yield the same labeling as the methods used for \emph{Hubble} and \emph{Spitzer} \citep{White2009}. 

For our accessibility analysis we use the entirety of the \emph{Chandra} and \emph{Spitzer} archives up until 2018 and 2017 where the data are currently complete, respectively. For \emph{Hubble} (launched in 1990), we use bibliographic data between 2000 and 2017 to cover a comparable range in time to the two newer missions. We examine data in the “archival” and “guest observer” categories, ignoring the “both” category for simplicity.

For each publication we use the NASA ADS to determine two additional parameters. The first is the primary institution of the first author. In the case where there are multiple institutions listed we only use the first institution listed for the first author. We use the canonical names of these institutions as reconciled by ADS, so that different spellings or nomenclatures used by the authors are not recorded as separate institutions. Additionally, for each of these institutions we record the country in which they are situated. Thus, for each archival and guest observer publication from each of the three missions we record a first author institution and a first author institution country. Only a few percent of each bibliography does not have an assigned institution or country; these data are excluded from the analysis.

As an additional resource, we use the gross domestic product per capita information from the World Bank online database for each country. 

For our update to \citet{White2009} and the scientific productivity of archives, we use the entire telescope bibliographies from \emph{Chandra}, \emph{Hubble}, and \emph{Spitzer} as well as the NASA/IPAC Extragalactic Database (NED) citations page\footnote{https://ned.ipac.caltech.edu/Acknowledgments/}. For \emph{Chandra}, \emph{Hubble}, and \emph{Spitzer} we use the “both” category in our analysis. 

\section{Results}
\begin{figure*}[ht!]
\includegraphics[width=\textwidth]{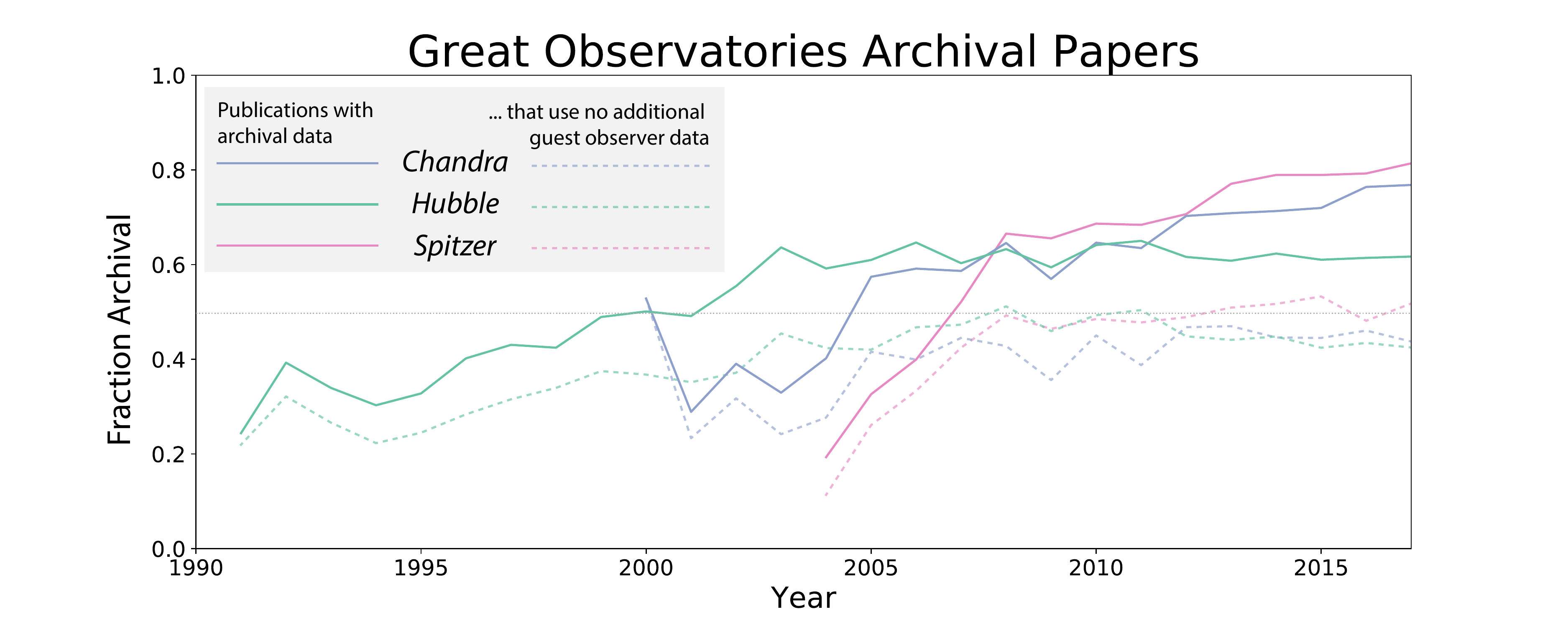}
\caption{Archival usage of \emph{Spitzer}, \emph{Chandra}, and \emph{Hubble} data. The dashed line represents the fraction of papers that use only archival data from the specified mission. The solid line represents the fraction of papers that use archival data from the specified mission as well as guest observer data.\label{arch_prod}}
\end{figure*}

\begin{figure*}[ht!]
\includegraphics[width=\textwidth]{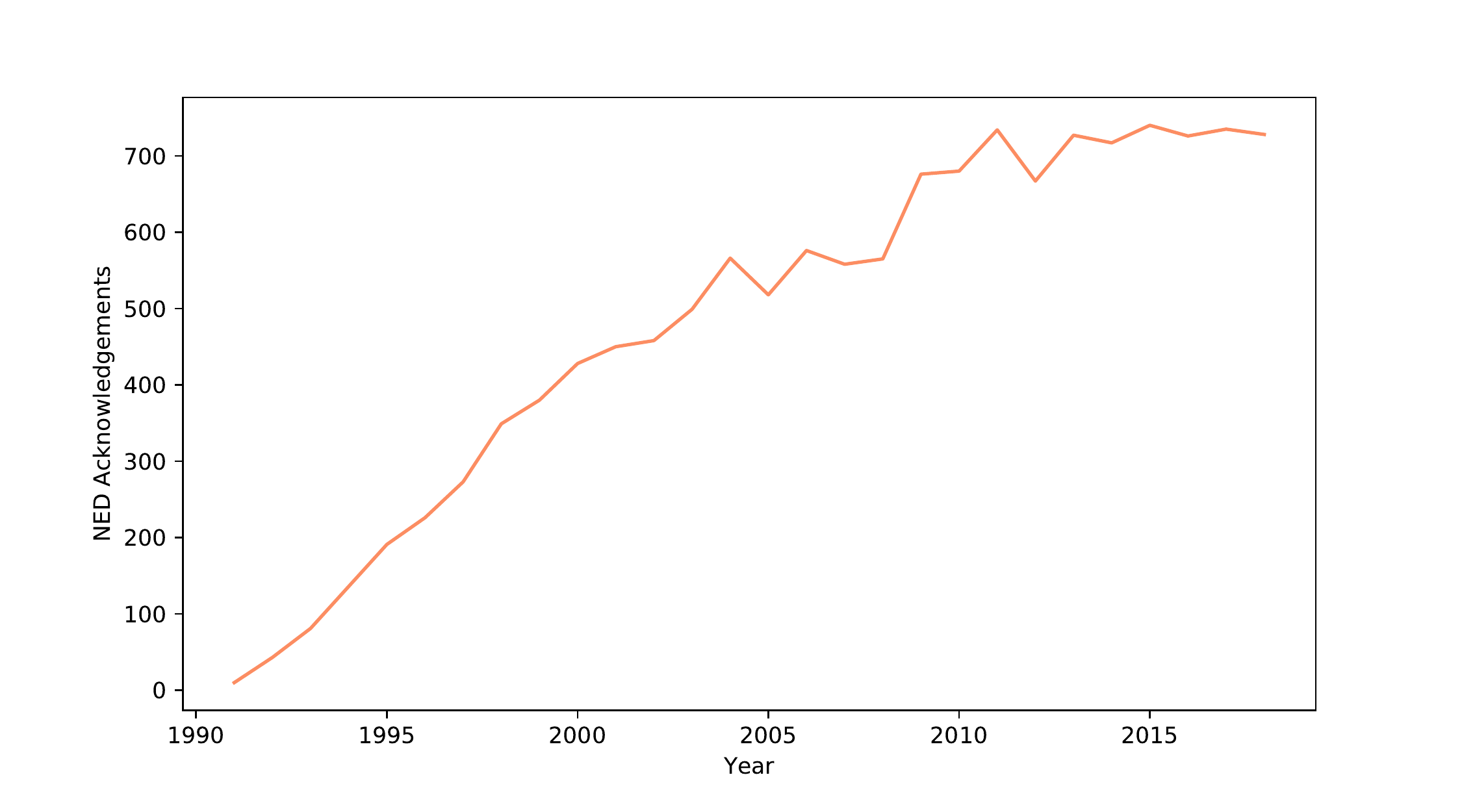}
\caption{Acknowledgments to the NASA/IPAC Extragalactic Database per year in the 
literature vs. time. Citations to NED represent entirely archival use of data, and they have grown to an average of approximately 730 peer-reviewed publications per year.
Bibliographic details are available at https://ned.ipac.caltech.edu/Acknowledgments/.
\label{ned}}
\end{figure*}

In our update to \citet{White2009}, we chart the fraction of all publications that are archival, either including or excluding the both category, shown in Figure \ref{arch_prod}. We also chart the total acknowledgements to NED in Figure \ref{ned}.

\begin{figure*}[ht!]
\includegraphics[width=\textwidth]{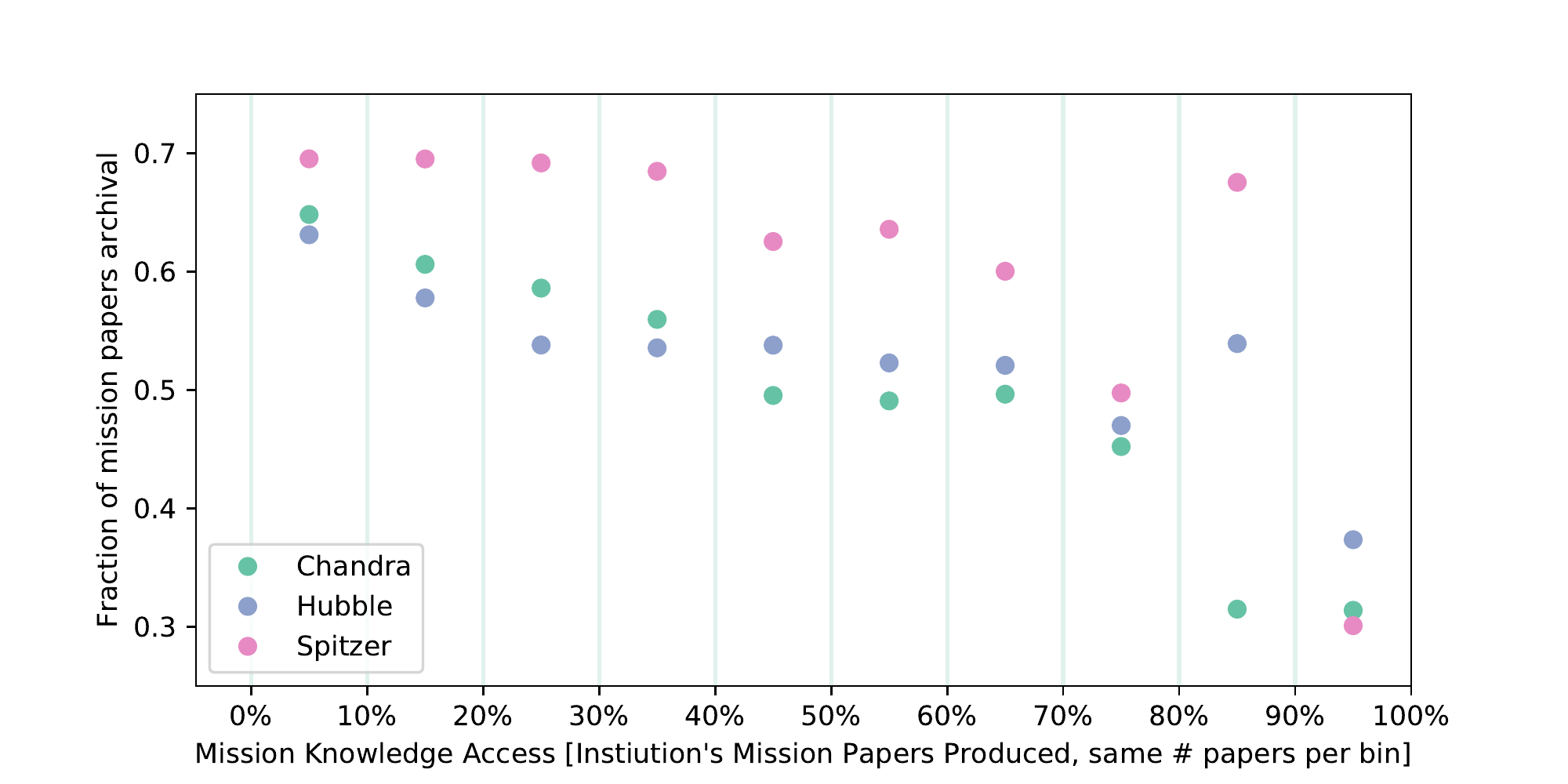}
\caption{We bin all archival and guest observer publications by each mission by their ``mission knowledge access'' score. The mission knowledge access for a publication is how many mission publications were published over the period that share the first author institution with the publication. Each bin here contains roughly the same number of papers for each mission. \label{mka}}
\end{figure*}

In our first scientific accessibility analysis we construct a parameter we call the “mission knowledge access”. We mark each publication from each mission by how many total publications from that mission have come from the first author’s institution over the period recorded. For example, if the first author of a \emph{Hubble} publication is from University of California, Irvine, we would give that publication a “mission knowledge access” score of 30, since 30 first authors published  \emph{Hubble} papers from University of California, Irvine over this period. We argue that first authors at institutions that have published few \emph{Hubble}-based publications will tend to have less access to local \emph{Hubble} knowledge, similarly for \emph{Chandra} and \emph{Spitzer}. We then take all publications from each mission we consider and bin it in deciles by mission knowledge access; each of the 10 bins has roughly the same number of publications. There are many more different first author institutions in lower deciles. For each bin we compute the fraction of all the publications in that bin that are archival. The results are shown in Figure \ref{mka}.

\begin{figure*}[ht!]
\begin{center}
\includegraphics{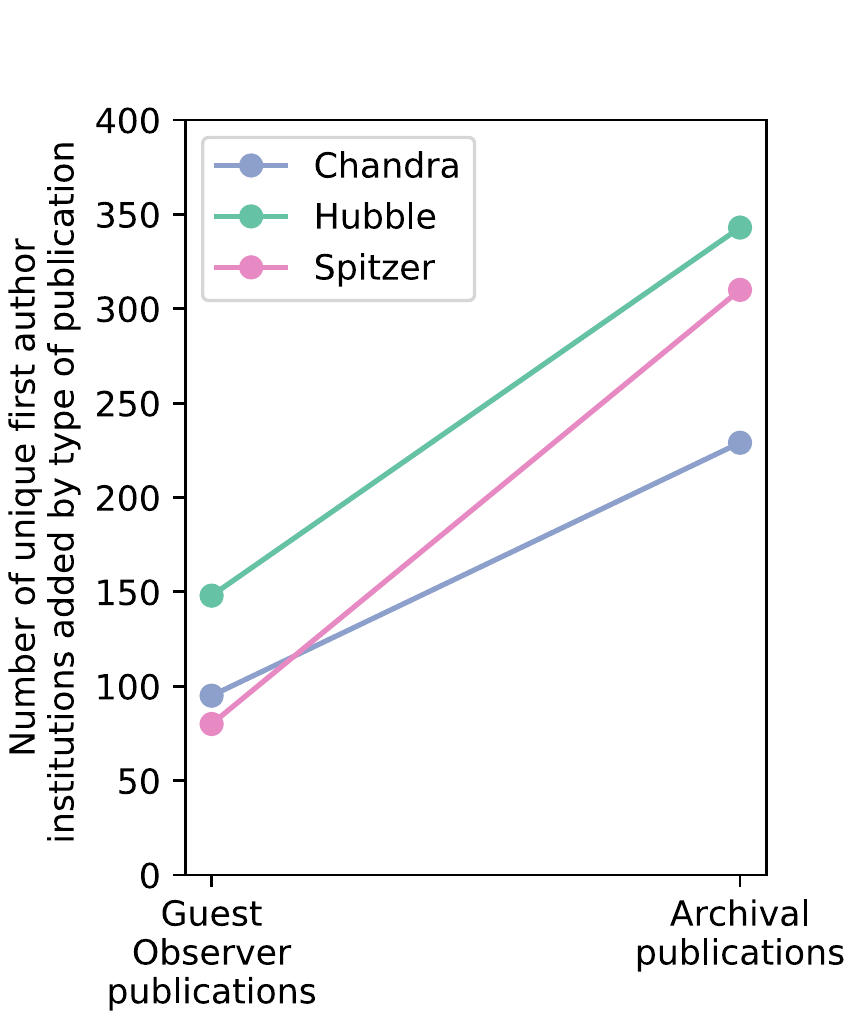}
\caption{The number of unique first author institutions added to the mission bibliography by guest observer publications and archival publications. \label{added_inst}}
\end{center}
\end{figure*}

We can also look at how important archival and guest observer publications are for adding unique institutions to the list of all institutions that have published using a given mission’s data. We find that archival publications add between 2 and 4 times more new institutions to a mission’s bibliographic base of first author institutions than guest observer publications (see Figure \ref{added_inst}).

\begin{figure*}[ht!]
\begin{center}
\includegraphics[width=0.8\textwidth]{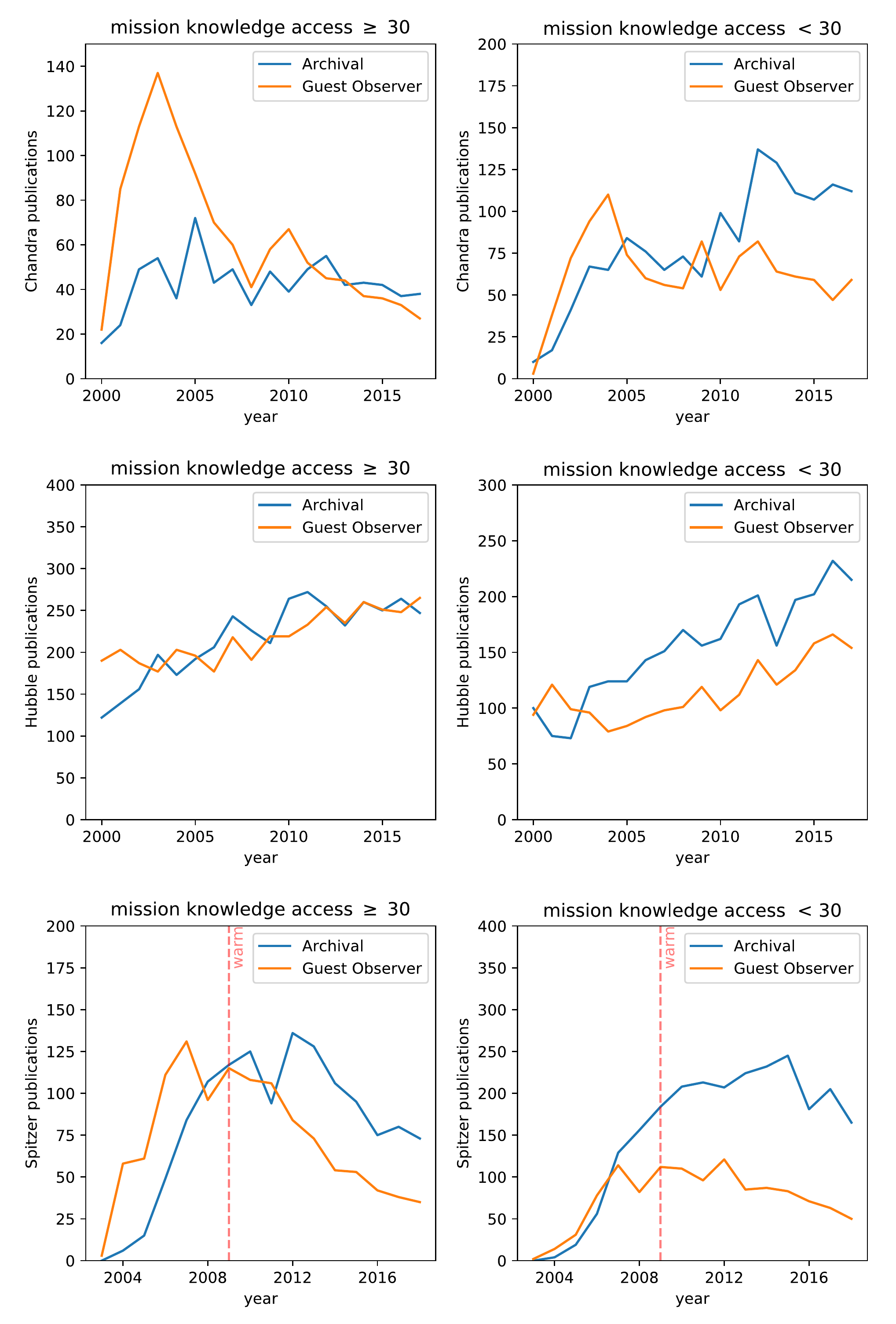}
\caption{The number Guest Observer and Archival publications for each mission split by number of publications from the same first author institution (``mission knowledge access''), plotted by year. \label{inst_pub_by_time}}
\end{center}
\end{figure*}

These data can also be split more coarsely by mission knowledge access and examined in time. In our second analysis we break up the information into two bins: publications from first-author institutions with fewer than 30 papers with mission data over the time period, and those with greater than or equal to 30. We then chart these against date for the three missions in Figure \ref{inst_pub_by_time}.

We can also assign a first author primary institution country to each publication. These countries each have a 2017 gross domestic product (GDP) per capita as found in the world bank database. GDP per capita is a rough instrument, but it does provide a universal proxy for the resources available to a researcher in a country. \citet{Kurtz2005} found that the basic research per capita in a country scales as (GDP per capita) squared.

\begin{figure*}[ht!]
\begin{center}
\includegraphics[width=\textwidth]{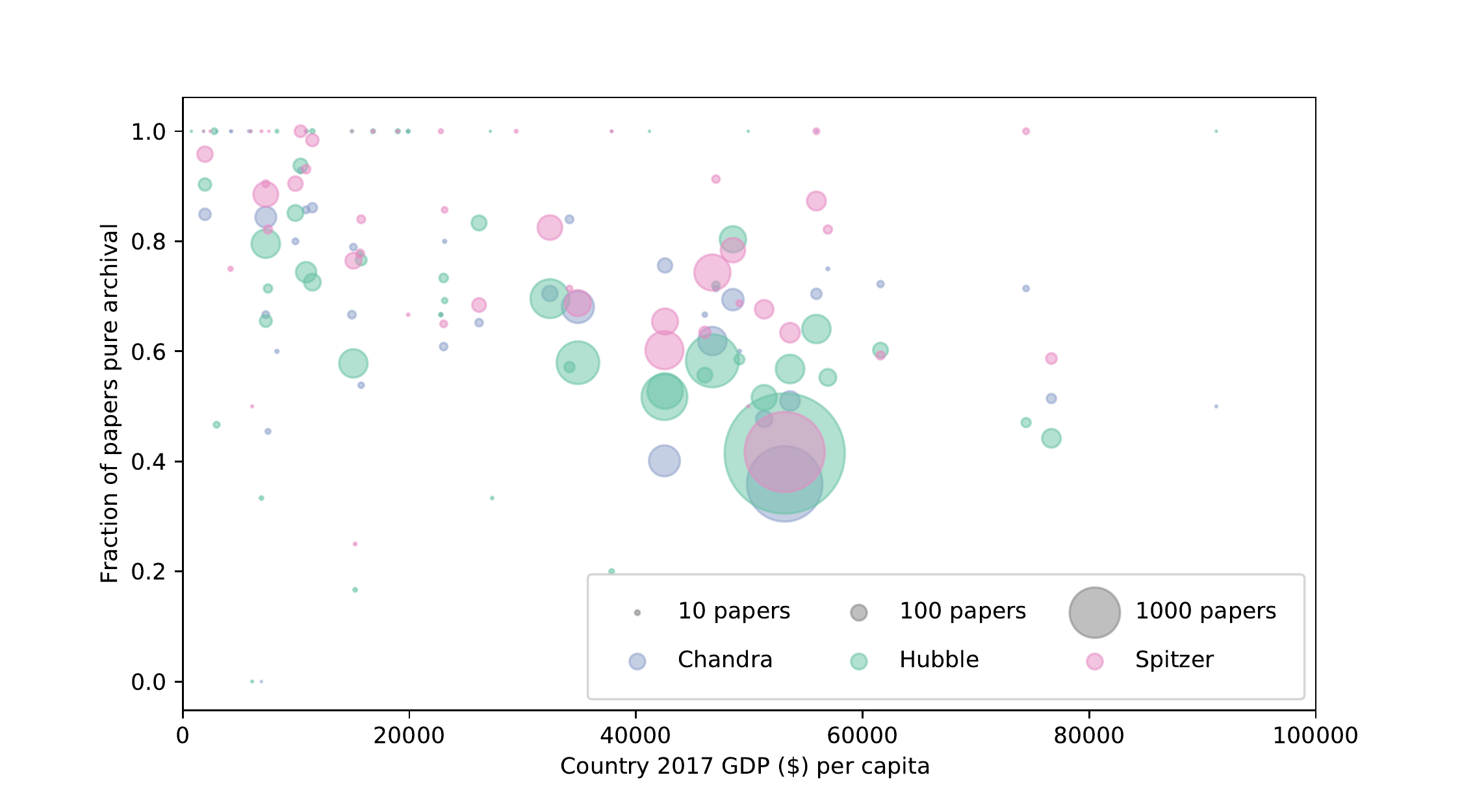}
\caption{The fraction of archival publications for each country, charted against GDP per capita. Larger dots represent more publications. \emph{Chandra}, \emph{Hubble}, and \emph{Spitzer} have 45, 55, and 51 countries represented in this analysis, respectively. \label{country_ARGO}}
\end{center}
\end{figure*}

\begin{figure*}[ht!]
\begin{center}
\includegraphics[width=0.8\textwidth]{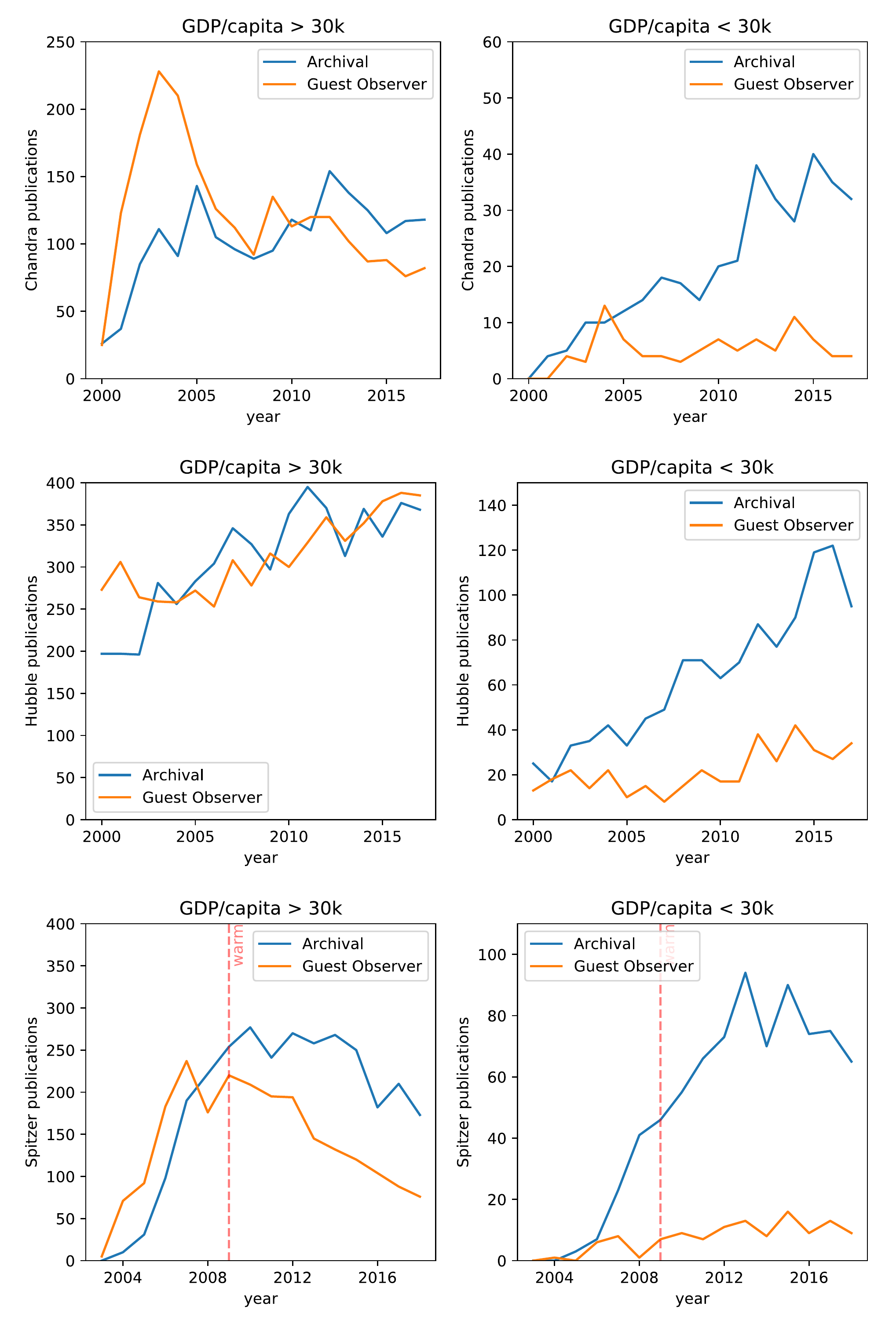}
\caption{The number Guest Observer and Archival publications for each mission split by GDP per capita of the first author institution country), plotted by year. \label{country_pub_by_time}}
\end{center}
\end{figure*}

We can perform the same time-domain analysis as we have for institutions, by splitting all publications into those whose first author institution country 2017 GDP per capita is less or more than \$30k in 2017 (Figure \ref{country_pub_by_time}).

\section{Analysis \& Discussion}

Figures \ref{arch_prod} and \ref{ned} have confirmed that the results of \citet{White2009} remain true. If anything, the trend has continued to rise, with more than 60\% of mission publications using some data from programs that share no co-Is with the publication authors. This trend seems quite strong and consistent over the decade. Similarly, NED publications, which are all archival in this sense, have continued to rise over the decade since \citet{White2009}.

Our analysis of publications split by first author institution shows a very strong and clear trend. Authors from institutions that have less “mission knowledge access”, proxied here by the number of total publications using mission data from that institution, much more often produce archival publications. Figure \ref{mka} shows that, roughly, bottom decile institutions in “mission knowledge access” publish 2/3 archival papers, while top decile institutions publish 1/3 archival papers.

The similar trends when splitting by first author institution country GDP per capita couldn’t be more stark, shown in Figure \ref{country_ARGO}. Countries with low GDP per capita publish 80-90\% archival publications, while countries with higher GDP per capita tend to cluster near 50\% archival publications. We note that \emph{Chandra}, \emph{Spitzer}, and \emph{Hubble} all have policies such that anyone in the world can propose to observe with these instruments. Further, NASA provides funding for both Guest Observer and Archival studies on all of these missions, either through the mission itself or through the Astrophysics Data Analysis Program only to US-based researchers. This means there is no specific funding or telescope access bias to our results.

Trends in time also show clearly how archival publications are growing in places with less access. In Figures \ref{inst_pub_by_time} and \ref{country_pub_by_time} we see that for institutions with less “mission knowledge access” and countries with less resources, archival publications have grown dramatically over the last ~15 years, much more so than non-archival publications. Various other trends are also visible, including the initial overall burst of publications (archival and guest observer) with the start of \emph{Chandra} and \emph{Spitzer} and the effect of the reduced observational capabilities of the \emph{Spitzer} warm mission.

These trends allow us to make a very clear statement: Archives are crucially important not just to maximize scientific productivity but also to maximize scientific accessibility. Scientists from a much more diverse set of institutions and countries publish using archival data, and thus archives dramatically increase the overall societal impact of our missions. Archives help enhance and spread the scientific value of our missions across institutions and across the world. Further, they help unite a broader group of researchers in the scientific process by providing common open data to the entire astronomical community.

\section{Recommendations}
 
\subsection{Follow the recommendations of White et al. 2009}

The results of \citet{White2009} still hold, and thus their ongoing recommendations are as important now as they were then. They are, by topline:
\begin{enumerate}

\item{Ensure sufficient funding and support for researchers making use of the archives}
\item{Ensure that current and future observatories and missions make their data available in science-ready form and with appropriate documentation}
\item{The fact that a good fraction of all science from missions will ultimately be archival should be taken into consideration in all phases of mission/observatory design and operations. The archival data products are, in the long term, as important as the PI science programs.}

\end{enumerate}
\subsection{Missions should prioritize archives to maximize community support}

To maximize the ongoing support of the astronomical community, space-based and ground-based facilities should prioritize well-resourced archives that hold science-ready data products. A well supported archive dramatically broadens the community that uses a mission and therefore broadens the community that will speak up to support the mission. There is always a temptation to “raid” the archive budget in mission planning in order to get the mission off the ground, and certainly there is no value in an archive without the mission itself. This said, we argue the last “bells and whistles” added to a mission to suit a very small number of observers are dramatically less important for the scientific productivity and accessibility of a mission than the archive funded by the equivalent resources. Even if missions are purely cynical, and uninterested in scientific accessibility for its own sake, the dramatically expanded base of users furnished by an archive will be a critical community for e.g. voicing broad support for extending mission lifetime.

\subsection{Archives must reach out to support a broader community and for input}

All the archives discussed here perform broad user surveys, and use this information to help inform decisions. This said, archive users’ groups tend to draw on elite users from large, US-based institutions with significant access to resources. These users will have trouble making recommendations that help the use cases of the typical archive user. We recommend here that archives use their bibliographies to find user’s group members from less well-known places that have less resource access. Further, we recommend that archives reach out to small institutions, both in the US and around the world, to help potential users make connections with the available data and services. We suggest that this could be accomplished digitally and through travel. To support the latter, we suggest more robust funding for travel than exists currently for archives.

\subsection{Recommendations apply from small telescopes to large surveys}

While our analysis has focused on data from NASA flagship observatories, there is no doubt that archives from smaller missions and ground-based observatories can have similar impacts on scientific diversity and productivity.  NASA has adopted a forward-looking policy that supports archives for all explorer-class and larger missions.  The resulting impact for some of these small missions, including Kepler and TESS, has been astonishing: over the past year, downloads of TESS data from MAST have exceeded even the volume of \emph{Hubble} data downloaded! Ground-based observatories have historically not had the funding required to create archives populated with calibrated, science-ready data products, but this has seen some improvements over the past decade at the U.S. national observatories.  We recommend that NSF continue to expand funding to support active, long-lived archives for its observatories.

Archival data access is even more important for survey projects, which are designed to generate a large uniform dataset.  The Sloan Digital Sky Survey project built a high-quality database that set the standard for accessibility and has had a huge scientific impact in astronomy. (ADS includes more than 17,000 papers that mention SDSS in their abstracts.) The success of SDSS has certainly influenced other projects.  The LSST project has built on the SDSS experience and has committed a significant fraction of its resources to data processing and the archive.  The decision of the \emph{Gaia} project to create a powerful catalog interface to accompany the early release of its data has led to an explosion of interest in the \emph{Gaia} products and has surely greatly expanded the community of users.  Future U.S. projects that are survey-oriented (e.g., WFIRST) should make the creation of a powerful, highly functional archive a top priority among the mission goals.

\section{Acknowledgements}
We thank Elizabeth Fraser and Jill Lagerstrom for help with telescope bibliographies.


\begin{thebibliography}{}

\bibitem[{Kurtz} et~al., 2005]{Kurtz2005}
{Kurtz}, M.~J., {Eichhorn}, G., {Accomazzi}, A., {Grant}, C.~S., {Demleitner},
  M., and {Murray}, S.~S. (2005).
\newblock {Worldwide Use and Impact of the NASA Astrophysics Data System
  Digital Library}.
\newblock {\em Journal of the American Society for Information Science and
  Technology}, 56:36.

\bibitem[{White} et~al., 2009]{White2009}
{White}, R.~L., {Accomazzi}, A., {Berriman}, G.~B., {Fabbiano}, G., {Madore},
  B.~F., {Mazzarella}, J.~M., {Rots}, A., {Smale}, A.~P., {Storrie-Lombardi},
  L., and {Winkelman}, S. (2009).
\newblock {The High Impact of Astronomical Data Archives}.
\newblock In {\em astro2010: The Astronomy and Astrophysics Decadal Survey},
  volume 2010, page P64.

\end{thebibliography}
\end{document}